\documentclass[aps,twocolumn,amsmath,amssymb, superscriptaddress, nofootinbib]{revtex4}
\usepackage[pdftex]{graphicx}
 \usepackage{amsmath}
\usepackage[T1]{fontenc}
\usepackage{amssymb}
\usepackage{amsfonts}
\usepackage{bm}
 \usepackage{amsmath} 
  \usepackage{flushend}
\usepackage{color}
\usepackage{lipsum}
\usepackage{amsfonts} 
\usepackage{amssymb, mathrsfs}
\usepackage{braket}
\usepackage{graphicx} 
\usepackage{subfigure}
\usepackage{bbm}
\def\beq{\begin{equation}}
\def\eeq{\end{equation}}
\def\bsp{\begin{split}}
\def\esp{\end{split}}
\def\bea{\begin{eqnarray}}
\def\eea{\end{eqnarray}}
\def\ba{\begin{array}}
\def\ea{\end{array}}

\def\dg{\dagger}

\def\lb{\left(}
\def\rb{\right)}

\def\l.{\left.}
\def\r.{\right.}

\def\ra{\rangle}
\def\la{\langle}

\def\bo{{\vec k}}

\begin{document}

\date{\today}
\title{Photoinduced topologically trivial magnons with finite thermal Hall effect}
\author{S. A. Owerre}
\affiliation{Perimeter Institute for Theoretical Physics, 31 Caroline St. N., Waterloo, Ontario N2L 2Y5, Canada.}

\begin{abstract}
In two-dimensional (2D) insulating magnets, the thermal Hall effect of magnons is believed to be a consequence of topological magnon insulator with separated magnon bands and a well-defined Chern number. Due to broken time-reversal symmetry the thermal Hall effect vanishes in Dirac magnons.   In this paper, we show that periodically driven  semi-Dirac magnon in 2D insulating honeycomb ferromagnet results in a photoinduced Dirac magnon at the topological phase transition between a photoinduced topological and trivial magnon insulator. Remarkably,  the photoinduced Dirac magnon and the photoinduced trivial magnon insulator possess a nonzero Berry curvature and exhibit a finite thermal Hall effect.  These  intriguing properties of periodically driven 2D insulating magnets originate from the bosonic nature of magnons. Hence, they are not expected to exist in 2D electronic Floquet systems. 
\end{abstract}
\maketitle

\section{Introduction}
Magnons in 2D insulating honeycomb ferromagnets have recently garnered considerable attention \cite{mag,sol,kkim,per,boy, yago,ruc, pie}, due to their similarity to electrons in graphene \cite{fdm, cas, kane}. Moreover, the recent experimental realization of 2D honeycomb ferromagnet in CrI$_3$ and other materials \cite{Huang, Gong} has provided a promising  possibility for realizing the magnonic analog of graphene.  The Dzyaloshinskii-Moriya (DM) interaction \cite{dm,dm2}, which can  be allowed  in materials with no inversion center is  forbidden by symmetry in most 2D insulating honeycomb ferromagnets. Therefore, the magnon dispersions of 2D insulating  honeycomb ferromagnet essentially possess  Dirac magnon points, reminiscent of Dirac fermions in graphene. However, as magnons are charge-neutral spin-1 bosonic quasiparticles, the  Dirac magnons, although very similar to Dirac fermions, are expected to exhibit different transport properties due to their quantum statistics. This discrepancy  has been overlooked in previous studies \cite{mag,sol,kkim,per,boy, yago,ruc, pie}. 

Indeed, magnons in the insulating magnets are simply magnetic dipole moments hopping on the lattice.  Therefore, in the presence of an electromagnetic field they accumulate the Aharonov-Casher phase \cite{aha}, similar to the  Aharonov-Bohm phase \cite{aha1} accumulated by charge particles in a perpendicular magnetic field. In recent years, the physics of magnons in the presence of a time-independent  electric field have been explored \cite{loss, ahaz, ahat, tak4},  and  was recently shown to realize magnonic Landau levels \cite{mei}. Quite distinctively, magnons hopping in the presence of a time-dependent oscillating electric field also lead to intriguing features. In particular, the present author has proposed that hopping magnons in the background of a time-dependent oscillating electric field  generate the magnonic Floquet topological insulator \cite{owe1}, with similar properties to  electronic Floquet topological insulator \cite{foot3, foot4, fot1, jot, eck1, foot5, eck2, eck3, tp6, tp7, tp8}. In the magnonic Floquet  topological insulator, an explicit DM interaction or scalar spin chirality is induced by circular-polarized electric field, which breaks the time-reversal symmetry of Dirac magnons.   
 
 In this paper, we study another intriguing property of periodically driven 2D insulating honeycomb ferromagnets.  We study the driven semi-Dirac  magnons with linear dispersion along the $k_y$ momentum direction and quadratic dispersion along the $k_x$ momentum direction.  We show that the driven semi-Dirac magnon results in a  photoinduced  Dirac magnon at the topological phase boundary between the magnonic Floquet topological and trivial insulators.   
 
 Usually, the thermal Hall effect in 2D insulating magnets is solely associated with topological magnon insulators with well-separated magnon bands and well-defined Chern numbers \cite{thm, thm1, thm2, thm3, thm4, thm5, thm6, sol1}. Thus, it is expected to vanish in the undriven gapless Dirac magnon due to the presence of time-reversal symmetry.  In contrast to this general belief, we find that the photoinduced Dirac magnon at the topological critical point possesses a nonzero Berry curvature, which generates a finite thermal Hall effect even without a well-defined Chern number.  We also find that the thermal Hall effect persists in the  magnonic Floquet trivial insulator phase. These results are not expected to manifest in the electronic Floquet counterparts. Therefore, they are as a consequence of the bosonic nature of magnons, which can only be thermally excited at low temperatures.  
 
\section{Model} 
\subsection{Spin model}

We consider the Heisenberg spin Hamiltonian for 2D insulating honeycomb ferromagnets governed by 
\begin{align}
\mathcal H&=-\sum_{ \la ij\ra } J_{ij}{\vec S}_{i}\cdot{\vec S}_{j}-g\mu_B B\sum_i S_i^z,
\label{spinh}
\end{align}
where $J_{i j}=J$ along the diagonal bonds and $J_{ij}=J^\prime$ along the vertical bonds. The last term is an external magnetic field with $g$ and $\mu_B$ being the spin $g$ factor and the Bohr magneton respectively. The spin Hamiltonian in Eq.~\eqref{spinh} is believed to describe many monolayer 2D insulating honeycomb ferromagnets  \cite{Huang, foot2}.

\subsection{Bosonic tight-binding model}
\begin{figure}
\centering
\includegraphics[width=1\linewidth]{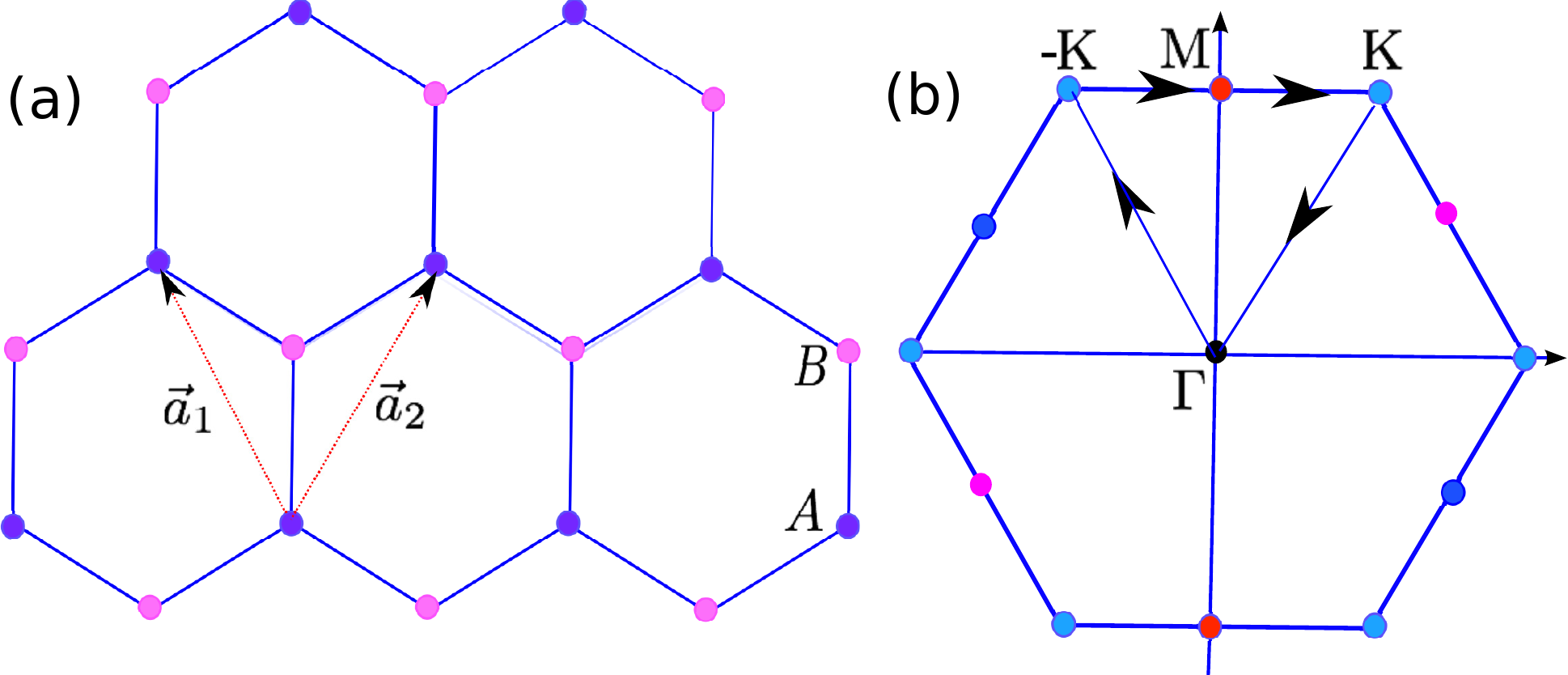}
\caption{Color online. (a) Schematic of the honeycomb lattice.   (b) The Brillouin zone (BZ) of the honeycomb lattice. Points with the same colour are related by symmetry.}
\label{lattice}
\end{figure}

 We describe the magnetic excitations of the spin Hamiltonian  Eq.~\eqref{spinh} by the Holstein Primakoff (HP)  transformation:  
 \begin{align}
  S_{i}^{ z}= S-a_{i}^\dagger a_{i},~S_{i}^+\approx \sqrt{2S}a_{i}=(S_{i}^-)^\dg,
 \end{align}
where $a_{i}^\dagger (a_{i})$ are the bosonic creation (annihilation) operators, and  $S^\pm_{i}= S^x_{i} \pm i S^y_{i}$ denote the spin raising and lowering  operators. By substituting the HP transformation into Eq.~\eqref{spinh}   and Fourier transforming   into  momentum space, we arrive at the free magnon Hamiltonian 
\begin{align}
\mathcal H=\sum_{\vec{k}} \psi_\bo^\dg\mathcal H({\vec{k}})\psi_\bo,
\end{align}
 with $\psi_\bo^\dg=(a_{\bo,A}^\dg,a_{\bo,B}^\dg)$.   
\begin{align}
\mathcal{H}(\bo)&=h_0{\bf 1} +h_x(\bo)\sigma_x + h_y(\bo)\sigma_y.
\label{honn}
\end{align}
where $\sigma_i~(i=x,y)$  are Pauli matrices with identity ${\bf 1}$. $h_0= t^\prime +2t+t_B$, with $~t(t^\prime)(t_B)=JS(J^\prime S)(g\mu_B B/S)$.  $h_x=\text{Re}[\rho(\bo)]$, $h_y=-\text{Im}[\rho(\bo)]$.    
\bea
\rho(\vec k)=-\sum_{\ell} t_\ell e^{ik_\ell},
\eea
where $t_1=t_2=t$, $t_3=t^\prime$. The momentum is $k_\ell = \vec k\cdot \vec a_\ell$ and the primitive vectors of the honeycomb lattice in Fig.~\ref{lattice}(a) are  $\vec a_{1,2}=a\big(\mp\sqrt{3}\hat x/2 + 3\hat y/2\big)$, and $\vec a_{3}=(0,0)$. Since the magnetic field simply rescales $h_0$, we will set $t_B=0$ in the following.  

\begin{figure}
\centering
\includegraphics[width=1\linewidth]{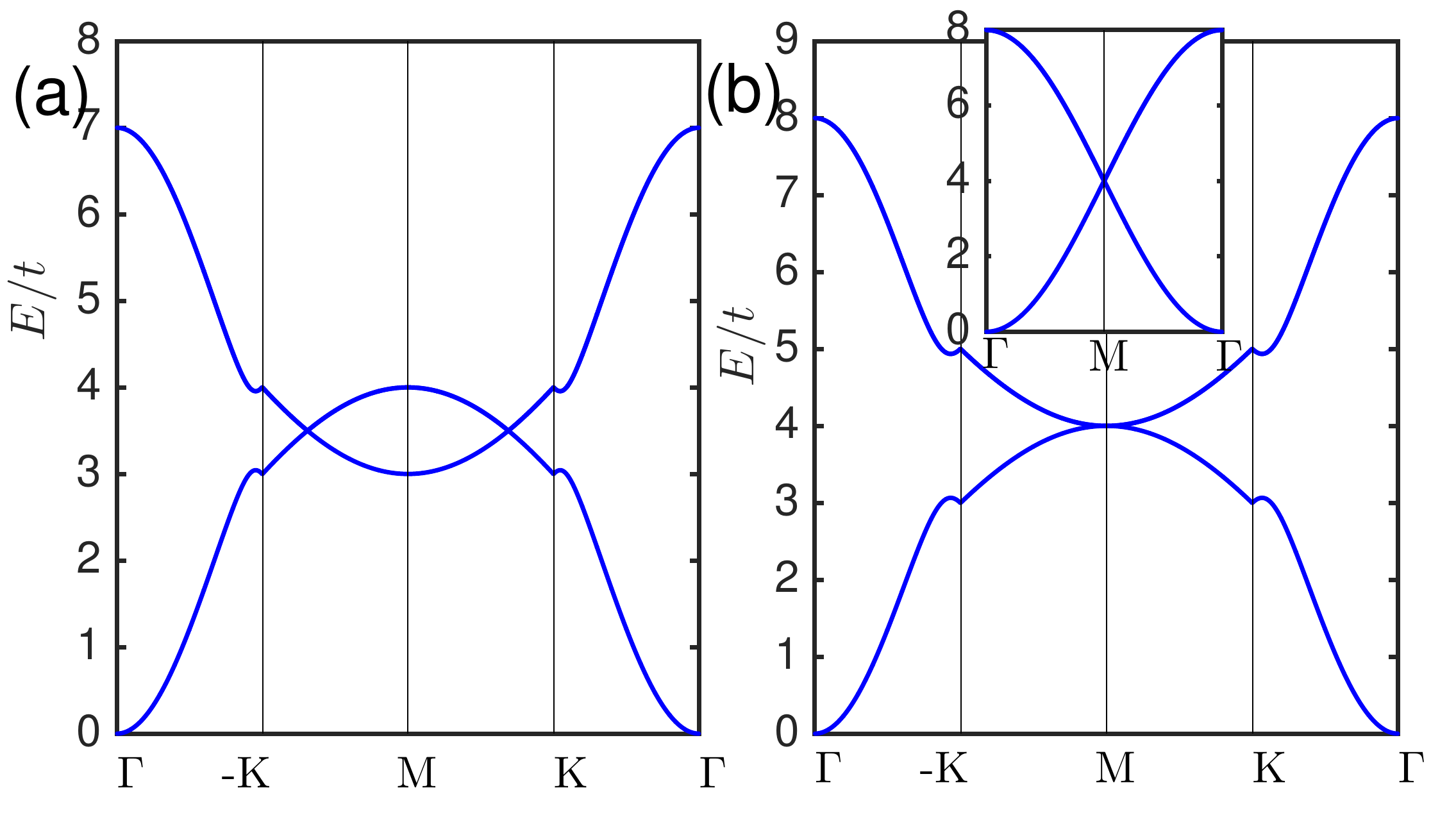}
\caption{Color online. The magnon bands of undriven 2D insulating honeycomb ferromagnet.  (a) Dirac magnons for  $t^\prime=1.5t$.  (b) Semi-Dirac magnons for $t^\prime=2t$. Inset shows the linear dispersion along the $k_y$ direction or ${\Gamma}$--{M}--${\Gamma}$ line.  }
\label{band1}
\end{figure}
 
\subsection{Dirac and Semi-Dirac magnons}
 For $t^\prime=t$, there are two Dirac magnon points at  $\pm K =(\pm 2\pi/3\sqrt{3},2\pi/3)$ in the Brillouin zone (BZ) Fig.~\ref{lattice}(b). They remain intact for $t^\prime<2t$, but move away from the   $\pm K$ point to  $\pm\tilde{K} =(\pm k_x^0,2\pi/3)$, where $k_x^0=\frac{2}{\sqrt 3}\arccos(t^\prime/2t)$ as shown in Fig.~\ref{band1}(a). Eventually, they emerge at $M$ $=(0, 2\pi/3)$ for $t^\prime=2t$ as shown in Fig.~\ref{band1}(b).  In this case, however, the magnon bands  are linear along the $k_y$ direction (see inset of (b)) and quadratic along the $k_x$ direction and form a semi-Dirac magnon band. Subsequently, a gap reopens for $t^\prime>2t$ and the system becomes a trivial magnon insulator.   We now obtain the effective Hamiltonian in the vicinity of the magnon band crossing points. 

Case ($i$): $t^\prime<2t$,  the effective Hamiltonian near $\bo=\pm\tilde{ K}$ is given by
\begin{align}
\mathcal{H}(\vec q+\eta\tilde{K})&=v_0{\bf 1}-(\eta v_x\sigma_x q_x +v_y\sigma_y q_y),
\label{ci}
\end{align}
where $\eta = \pm$, $v_0=h_0$, $v_x=\sqrt{3}t\sin(\sqrt{3}k_x^0/2)$ and $v_y=3t\cos(\sqrt{3}k_x^0/2)$.

Case ($ii$): $t^\prime= 2t$,  the effective Hamiltonian near $\bo=$ \text{ {M}} is given by
\begin{align}
\mathcal{H}(\vec q+M)&=v_0{\bf 1}-\frac{q_x^2}{2m}\sigma_x  -  \tilde v_t\sigma_y q_y,
\label{cii}
\end{align}
where $m=1/v_D$ and $\tilde v_t=2v_D$ with $v_D=3t/2$. This is the semi-Dirac magnon Hamiltonian which is our focus in this paper. 

\section{Floquet Magnons}

\subsection{Time-dependent Aharonov-Casher phase}
Magnons in 2D insulating magnets are simply hopping magnetic dipole moment  $g\mu_B\hat z$. Hence, in the presence  of  a time-periodic electric field $\vec{E}(\tau)$, they accumulate the  time-dependent Aharonov-Casher phase (see Appendix) 
\begin{align}
\theta_{ij}(\tau)=\mu_m\int_{\vec{r}_{i}}^{\vec{r}_{j}} \vec{E}(\tau)\times \hat{z}\cdot d{\vec \ell},
\end{align}
where $\mu_m = g\mu_B/\hbar c^2$, $\hbar = h/2\pi$ and $c$ are  the reduced Planck's constant and the speed of light respectively, and  ${\vec r}_i$ is the coordinate of the lattice at site $i$. The electric field obey the relation:  $\vec{E}(\tau)=-\partial \vec{A}(\tau)/\partial \tau$, where $ \vec{A}(\tau)$ is the time-periodic vector potential given by
\begin{align}
 &\vec A(\tau) = \big[A_x\sin(\omega\tau + \phi), A_y\cos(\omega\tau),0\big],
 \end{align}
where $A_{i}= E_{i}/\omega$ ($i = x, y$) are the amplitudes of the time-dependent  vector potential. Here $\omega$ is the frequency of the light and $\phi$ is the phase difference.  Hence, the electric field is given by
\begin{align}
\vec{E}(\tau)\times \hat z=[E_y\sin(\omega \tau),E_x\sin(\omega \tau+\phi),0]
\end{align}
where    $E_x$ and $E_y$ are the amplitudes of the time-periodic electric field along the $x$ and $y$ directions.  Note that the time-periodicity is given by $\vec{E}(\tau)= \vec{E}(\tau + T)$ where $T=2\pi/\omega$ is the period.  The  resulting time-dependent Hamiltonian $\mathcal H(\vec{k},\tau)$ is obtained by making  the  time-dependent Peierls substitution  ${\vec k}\to {\vec k} +\mu_m\big(\vec{E}(\tau)\times \hat{z}\big) $ into Eq.~\eqref{honn}.  Therefore, we can now study periodically driven  magnon systems in a same manner to driven electronic systems. 

\subsection{ Floquet-Bloch formalism}
 The Hamiltonian  can now be expanded  in Fourier space as  

\begin{align}
\mathcal{H}(\vec{k},\tau)= \sum_{n=-\infty}^{\infty} e^{in\omega \tau}\mathcal{H}_n(\vec{k}),
\end{align}
where  the Fourier component is given by
$\mathcal{H}_n(\vec{k})=\frac{1}{T}\int_{0}^T e^{-in\omega \tau}\mathcal{H}(\vec{k}, \tau) d\tau=\mathcal{H}_{-n}^\dg(\vec{k}).$ Its eigenvectors has the  form $\ket{\psi_\alpha(\vec{k}, \tau)}=e^{-i \epsilon_\alpha(\vec{k}) \tau}\ket{\xi_\alpha(\vec{k}, \tau)}$, where  $\ket{\xi_\alpha(\vec{k}, \tau)}=\sum_{n} e^{in\omega \tau}\ket{\xi_{\alpha}^n(\vec{k})}$ is the time-periodic Floquet-Bloch wave function of magnons and $\epsilon_\alpha(\vec{k})$ are the magnon quasienergies. The Floquet operator is defined as $\mathcal{H}^F(\vec{k},\tau)=\mathcal{H}(\vec{k},\tau)-i\partial_\tau$, which  leads  to the Floquet eigenvalue equation

\begin{align}
\sum_m [\mathcal H_{n-m}(\vec{ k}) + m\omega\delta_{n,m}]\xi_{\alpha}^m(\vec{k})= \epsilon_\alpha(\vec{k})\xi_{\alpha}^n(\vec{k}).
\end{align}
The time-dependent  Hamiltonian $\mathcal H({\vec k},\tau)$  is given by
\begin{align}
\mathcal H(\vec k,\tau)= \rho_0{\bf 1}+
\begin{pmatrix}
0&\rho(\bo,\tau)\\
\rho^*(\bo,\tau)&0\\
\end{pmatrix},
\label{tham1}
\end{align}
where $\rho_0=h_0$ and $\rho(\bo,\tau) =-\sum_{\ell}t_\ell e^{ik_\ell}e^{i\mu_m\vec {E}(\tau)\cdot \vec{\delta}_\ell}$. Here  $\vec \delta_1=a(\sqrt{3}/2, -1/2)$, $\vec \delta_2=-a(\sqrt{3}/2, 1/2)$, and $\vec \delta_3=a(0, 1)$ are the vectors of the nearest-neighbour bonds. The Fourier components of the Hamiltonian \eqref{tham1} are given by

\begin{align}
\mathcal H_q(\vec k)&=\rho_0\delta_{q,0}{\bf 1}+
\begin{pmatrix}
0&\rho_q(\bo)\\
\rho_{-q}^*(\bo)&0\\
\end{pmatrix},
\label{fham1}
\end{align}
where $\rho_q(\bo)=-\sum_\ell t_{\ell, q} e^{ik_\ell}$, and $q= n-m = -1,0,1$. The renormalized interactions  are  given by 

\begin{align}
& t_{1,q}=t\mathcal  J_{-q}(\mathcal E_{-})e^{-iq\Psi_{-}},~ t_{2,q}=t\mathcal J_q(\mathcal{E}_{+})e^{iq\Psi_{+}},\nonumber\\& t_{3,q}= t^\prime\mathcal J_{q}(\mathcal{E}_{x})e^{iq\phi},
\end{align}
where $\mathcal  J_n(x)$ is the Bessel function of order $n$.
\begin{align}
 &\mathcal{E}_{\pm}=\frac{1}{2}\sqrt{3\mathcal E_y^2+\mathcal E_x^2\pm 2\sqrt{3}\mathcal E_x \mathcal E_y\cos(\phi)},\\ &\Psi_{\pm}=\arctan\lb \frac{\mathcal E_x\sin(\phi)}{\sqrt{3}\mathcal E_y\pm \mathcal E_x\cos(\phi)}\rb.
 \end{align}
 
 In the  magnonic Floquet formalism, the light intensity  is characterized by the dimensionless quantity
\begin{align}
\mathcal E_i =\frac{g\mu_B E_i a}{\hbar c^2}.
\label{dims}
\end{align}
where $i = x,y$ and $a$ is the lattice constant. Since there is no frequency denominator in Eq.~\ref{dims}, the magnonic Floquet formalism is different from that of electronic Floquet formalism \cite{foot3, foot5}.

 \subsection{Magnonic  Floquet quasienergy bands}
 
We consider the  off-resonant limit $\omega\gg t,t^\prime$.  In this limit, the system can be described by an effective time-independent  Hamiltonian \cite{tp6,tp7,tp8}, which can be obtained perturbatively in $1/\omega$ expansion as 
\begin{align}
\mathcal H_{\text {eff}}(\vec k)&=\mathcal H_0(\vec k)-\frac{1}{\omega}\big( \big[\mathcal H_0(\vec k), \mathcal H_{-1}(\vec k)\big]-\big[\mathcal H_0(\vec k), \mathcal H_{1}(\vec k)\big]\nonumber\\&+\big[\mathcal H_{-1}(\vec k), \mathcal H_{1}(\vec k)\big]\big).
\label{effHam}
\end{align}

\begin{figure}
\centering
\includegraphics[width=1\linewidth]{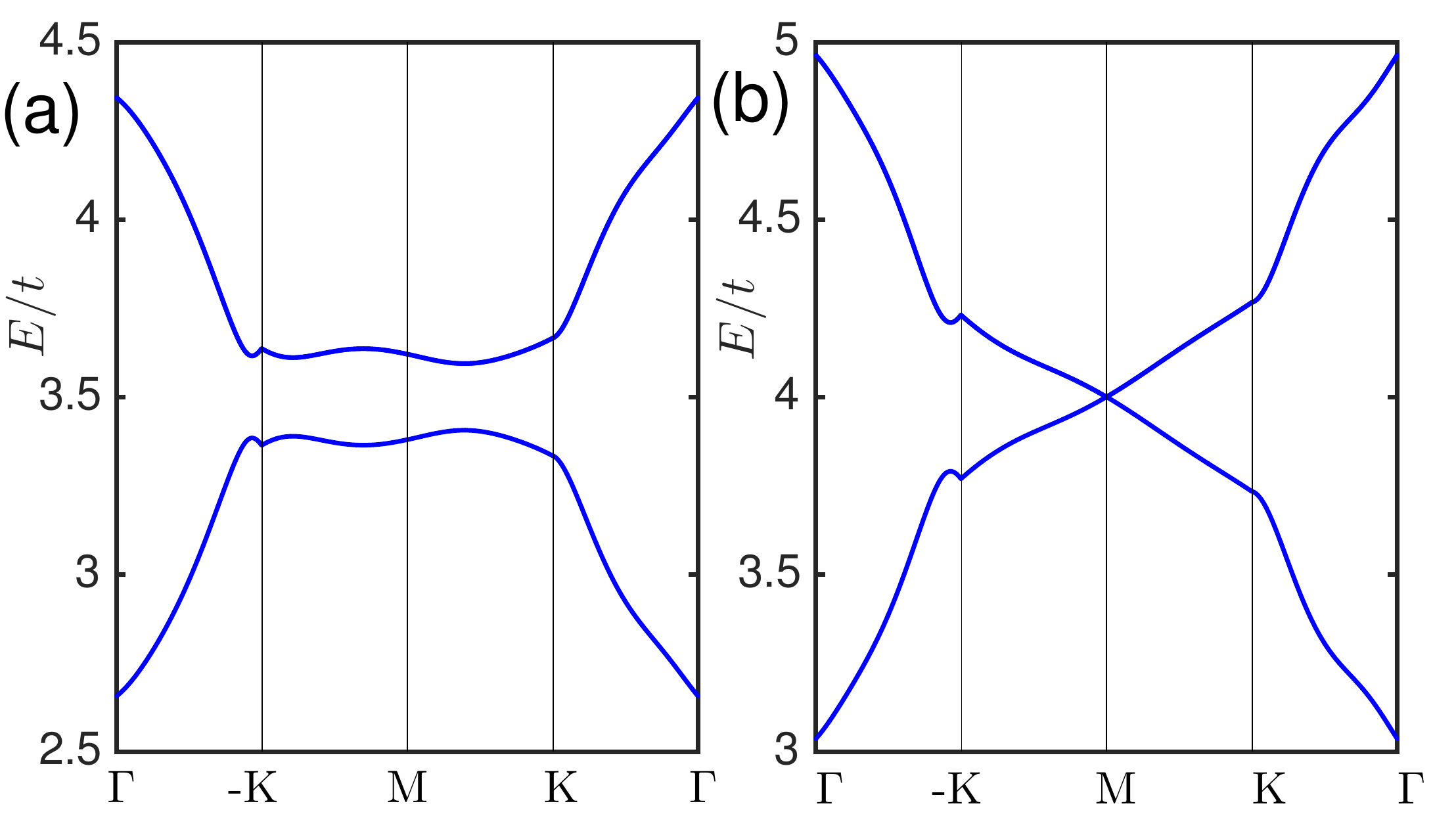}
\caption{Color online. Magnonic Floquet  quasienergy of periodically driven 2D insulating honeycomb ferromagnet. (a)  Magnonic Floquet topological insulator for $t^\prime=1.5t$. (b) Photoinduced Dirac magnons in periodically driven semi-Dirac magnons for $t^\prime=2t$. Here we use circularly-polarized electric field $\phi=\pi/2$, $\mathcal E_x= \mathcal E_y=2$ in units of $g\mu_B a/\hbar c^2$, and $\omega=10t$.}
\label{band2}
\end{figure}

In Fig.~\ref{band2}, we have shown the Floquet magnon quasienergy bands. Fig.~\ref{band2}(a) shows the magnonic Floquet topological insulator for $t^\prime<2t$ and Fig.~\ref{band2}(b) shows the photoinduced Dirac magnon for $t^\prime=2t$. We can see that driven semi-Dirac magnon remain gapless  at $t^\prime=2t$, due to the competition between $k_x$ and $k_y$ (cf. Eq.~\ref{cii}). Note that the Floquet magnon quasienergy bands are gapped in the regime $t^\prime>2t$ (not shown), which corresponds to a Floquet trivial magnon insulator. As we will show later, the  Floquet trivial magnon insulator for $t^\prime>2t$ and $\phi=\pi/2$  is different from the one induced by linearly-polarized light for $\phi=0$. It is crucial to note that  the photoinduced Dirac magnon in Fig.~\ref{band2}(b)  is reminiscent of 2D Dirac semimetals with SOC \cite{ste}. In fact, for $\phi=\pi/2$ and $\mathcal E_{x}=\mathcal E_y= \mathcal E$, the effective Hamiltonian of the photoinduced Dirac magnons near $\bo=  M$ is given by
\begin{align}
\mathcal{H}(\vec q+M)&=v_0{\bf 1}+v_x(\mathcal E,\omega)q_x\sigma_z -v_y(\mathcal E)\sigma_y q_y,
\label{pcii}
\end{align}
where $v_x(\mathcal E,\omega)$ and $v_y(\mathcal E)$ are the group velocities along the $x$ and $y$ directions respectively. 

\begin{figure}
\centering
\includegraphics[width=1\linewidth]{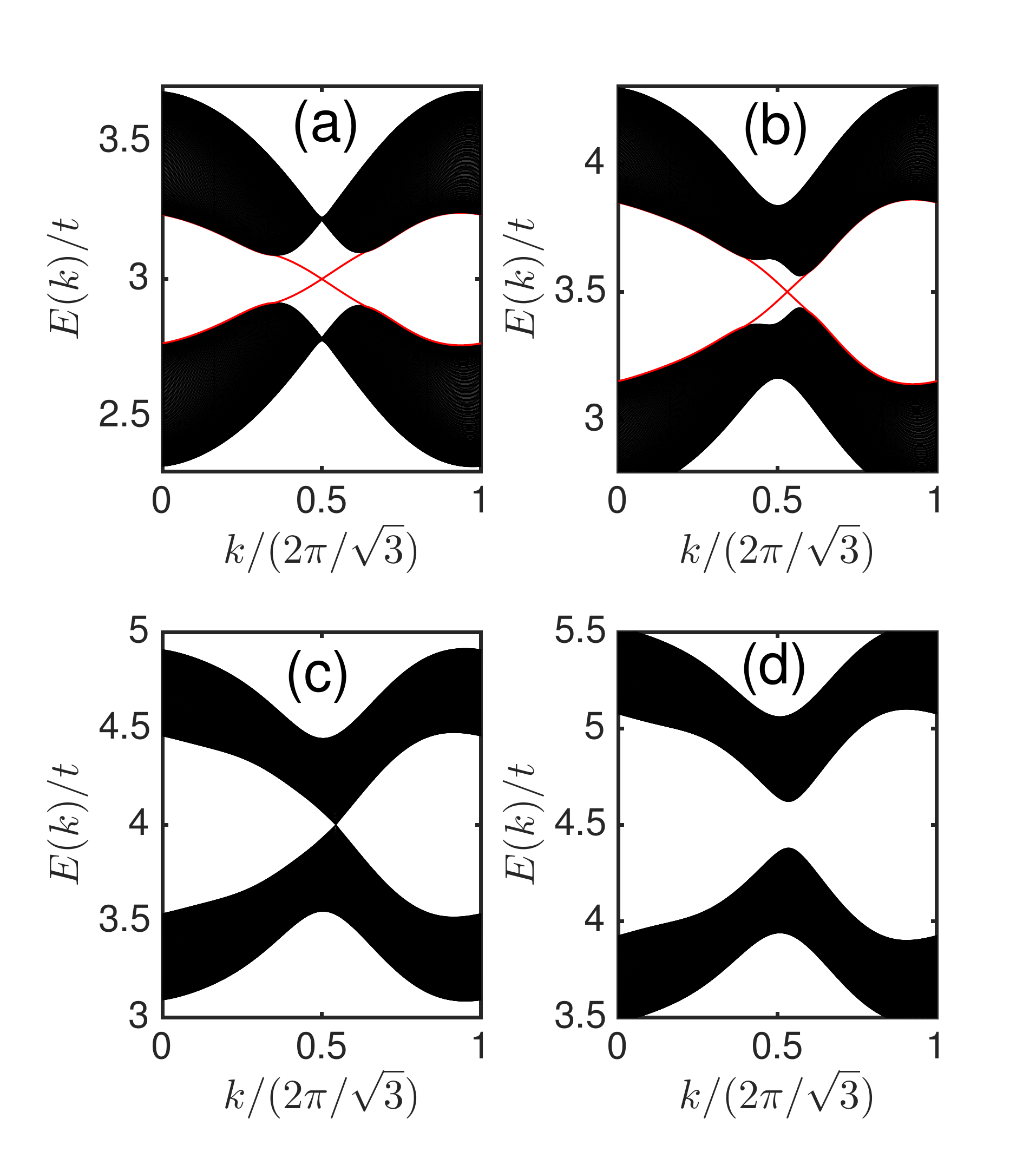}
\caption{Color online. Magnonic Floquet  quasienergy bands for a one-dimensional strip geometry in different phases.  (a) and (b) Magnonic Floquet topological insulators  for   $t^\prime=t$ and $t^\prime=1.5t$ respectively.  (c) Photoinduced Dirac magnon phase  for $t^\prime=2t$, (d) Magnonic Floquet trivial insulator for $t^\prime=2.5t$. Here  we use circularly-polarized electric field $\phi=\pi/2$, $\mathcal E_x= \mathcal E_y=2$ in units of $g\mu_B a/\hbar c^2$, and $\omega=10t$.}
\label{edge}
\end{figure}

We further show the Floquet  magnon quasienergy bands for a 1D strip geometry in Fig.~\ref{edge}. In the magnonic Floquet topological insulator phase (a) and (b), there are chiral gapless magnon edge modes traversing the bulk magnon gap. For the photoinduced Dirac magnons in (c), we can see that the bulk gap closes --- an indication of a topological phase transition to a magnonic Floquet trivial insulator  in (d), with no chiral magnon edge modes.


 \begin{figure*}
\centering
\includegraphics[width=1\linewidth]{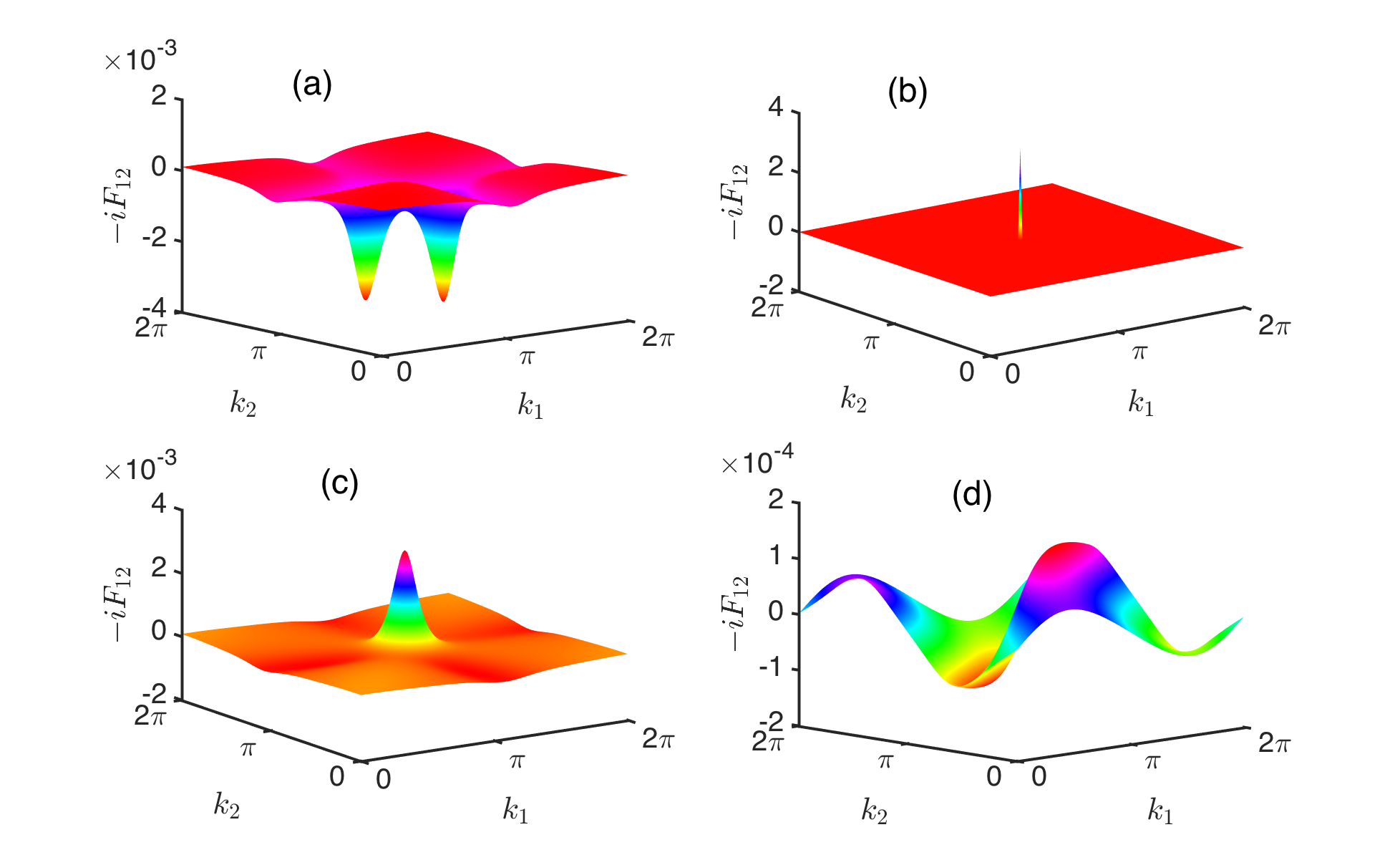}
\caption{Color online. Photoinduced Berry curvature distribution of the lowest Floquet magnon band for the (a) magnonic Floquet topological insulator induced  by circularly-polarized light $\phi=\pi/2$ for $t^\prime=1.5t$, (b) photoinduced Dirac magnon induced  by circularly-polarized light $\phi=\pi/2$ for $t^\prime=2t$, (c) magnonic Floquet trivial insulator induced by circularly-polarized light $\phi=\pi/2$  for $t^\prime=2.5t$, (d) magnonic Floquet trivial insulator induced by linearly-polarized light  $\phi=0$ for $t^\prime=1.5t$. The parameters in all plots are $\mathcal E_x= \mathcal E_y=2$ in units of $g\mu_B a/\hbar c^2$, and $\omega=10t$. }
\label{berry}
\end{figure*}

 \begin{figure}
\centering
\includegraphics[width=1\linewidth]{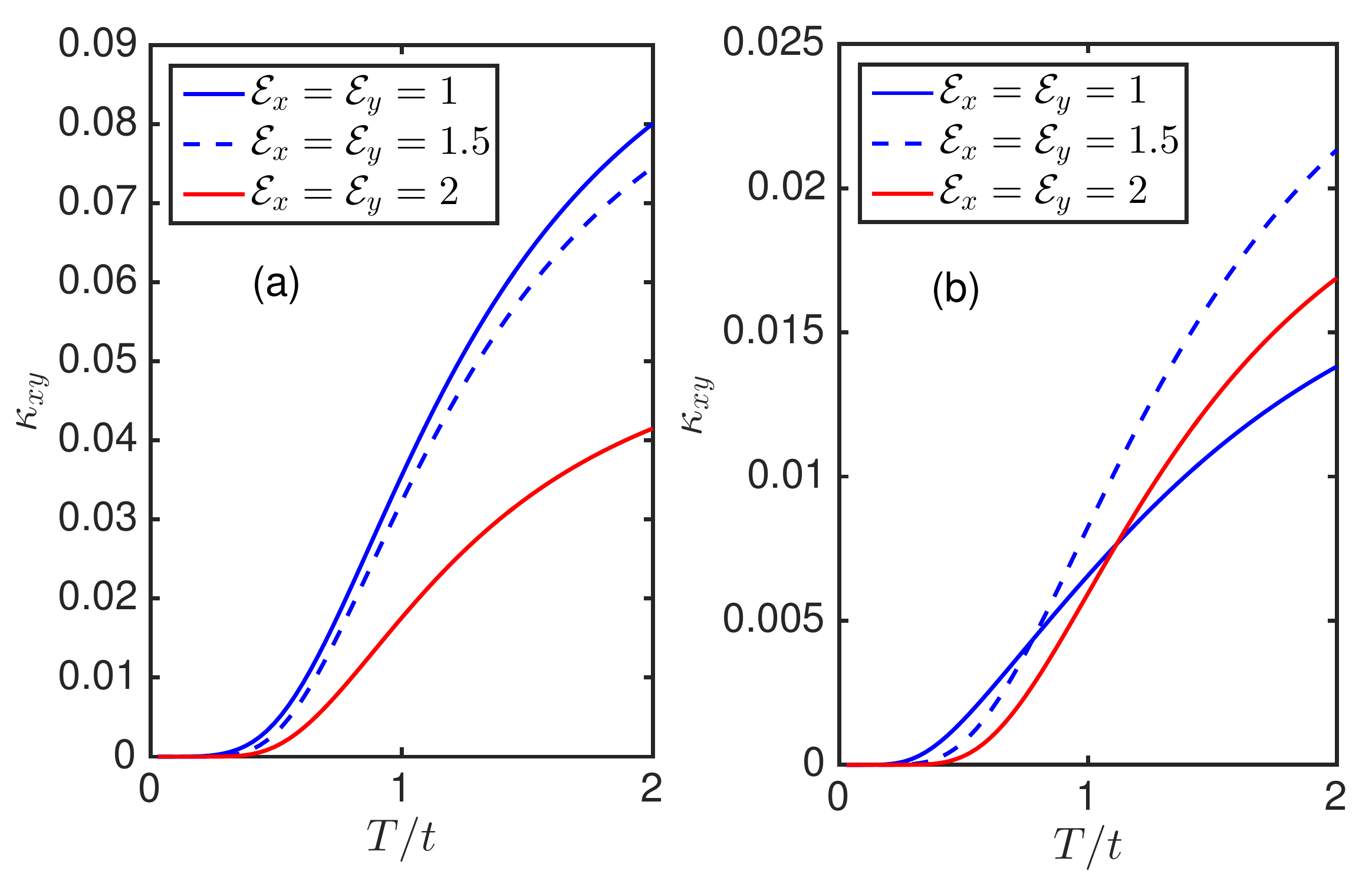}
\caption{Color online. Photoinduced thermal Hall conductivity for different light amplitudes in units of $g\mu_B a/\hbar c^2$. (a) Photoinduced Dirac magnon phase for $t^\prime=2t$.  (b) Photoinduced trivial  magnon insulator phase  for $t^\prime=2.5t$. We use circularly-polarized electric field $\phi=\pi/2$ and $\omega=10t$.}
\label{THE}
\end{figure}

 \subsection{Topological phase transiton}
To compute the  Berry curvature and the Chern number of the  magnon quasienergy bands, we implement the discretize Brillouin zone method proposed by T. Fukui et al. \cite{fuk}. In this method, the Berry curvature is of the form $\Omega(\vec k_\ell) \equiv F_{12}(\vec k_\ell)/i$, where $F_{12}(\vec k_\ell)$ is the lattice field strength, which is defined by U(1) links \cite{fuk}. The associated Chern number is defined as,
\begin{align}
\mathcal C=\frac{1}{2\pi i}\sum_\ell  F_{12}(\vec k_\ell).
\end{align}

  In Fig.~\ref{berry} we have shown the photoinduced Berry curvature distribution of the lowest magnon band.  As shown in Fig.~\ref{berry}(a), the photoinduced Berry curvature has two peaks in the magnonic Floquet topological insulator phase. They correspond to the photoinduced  gaps at the Dirac points. In Fig.~\ref{berry}(b) we can also see that the photoinduced Berry curvature vanishes everywhere except at the photoinduced Dirac magnon, where it develops a large peak. Interestingly,  the magnonic Floquet trivial  insulator induced by circularly-polarized light at $\phi = \pi/2$ and $t^\prime >2t$ has a different photoinduced Berry curvature distribution  from the one induced by linearly-polarized light at  $\phi = 0$. This can be seen in Figs.~\ref{berry} (c) and (d) respectively.   We find that the Chern number changes from $\mathcal C=\pm 1$ in the magnonic Floquet topological insulator phase ($t^\prime<2t$) to $\mathcal C=0$ in the magnonic Floquet trivial insulator phase ($t^\prime>2t$) for $\phi = \pi/2$ or $\phi = 0$. Therefore, the  photoinduced Dirac magnon phase at $t^\prime=2t$ is a topological critical point. 

\subsection{Thermal Hall effect}

The most interesting aspects of ferromagnetic topological magnons  is that they exhibit the  thermal Hall effect \cite{thm, thm1, thm2, thm3, thm4, thm5, thm6, sol1}. In ferromagnetic insulators, the  thermal Hall effect has only been studied in the topological magnon insulator phase, when the lowest (acoustic) magnon band is well separated  and carry a well-defined Chern number. For periodically driven magnon systems, we focus on the regime where the Bose distribution function is close to thermal equilibrium. In this regime, the same theoretical concept of the thermal Hall effect in undriven topological magnon systems \cite{thm, thm1, thm2, thm3, thm4, thm5, thm6, sol1}  can be applied to the driven magnon systems.  The transverse component  of the thermal Hall conductivity is given explicitly by \cite{thm5}
\begin{align}
\kappa_{xy}=-\frac{k_B^2 T}{V}\sum_{\vec k} \sum_{\pm} c_2\lb n_\alpha\rb\Omega_{\pm}(\vec k),
\label{thm}
\end{align}
where  $V$ is the volume of the system, $T$ and  $k_B$ are the temperature and the Boltzmann constant respectively.  
\bea
 n_\pm=n[ \epsilon_{\pm}(\vec k)]= \frac{1}{\exp\big({\epsilon_{\pm}(\vec k)/k_BT\big)}-1},
 \eea is the Bose distribution function close to thermal equilibrium. Here  $ c_2(x)=(1+x)\lb \ln \frac{1+x}{x}\rb^2-(\ln x)^2-2\text{Li}_2(-x)$, with $\text{Li}_2(x)$ being the  dilogarithm. The plus or minus sign refers to the two middle magnon quasienergies. 

 Evidently, the thermal Hall conductivity is simply the Berry curvature weighed by the $c_2$ function. Therefore,  its dominant contribution comes from the peaks of the Berry curvature (see Fig.~\ref{berry}). In addition, the thermal Hall transport is determined mainly by the acoustic (lower) magnon branch due to the bosonic nature of magnons. In Fig.~\ref{THE}(a) and (b), we have shown the trend of $\kappa_{xy}$ for different electric field amplitudes in the (a) photoinduced Dirac magnon phase for $t^\prime=2t$ and (b) in the photoinduced trivial  magnon insulator phase  for $t^\prime>2t$.  The finite thermal Hall conductivity in these topologically trivial phases can be attributed to the photoinduced  Berry curvatures in Figs.~\ref{berry}(b) and (c). We note that the thermal Hall conductivity vanishes for the  magnonic Floquet  trivial  insulator induced by linearly-polarized light $\phi =0$, because the integration  of the Berry curvature in Fig.~\ref{berry}(d) vanishes identically. This is expected as linearly-polarized light does not break time-reversal symmetry.

\section{Conclusion}
We have shown that photoinduced Dirac magnons at the topological phase transition can exhibit interesting transport property that is not expected in the undriven Dirac magnon systems, as well as in electronic systems. Furthermore, we have established   that   magnon edge modes and well-defined Chern number of magnon bands do not guarantee the presence or absence of the thermal Hall effect in 2D ferromagnetic insulators. In other words, the thermal Hall effect in 2D ferromagnetic insulators is not necessarily a consequence of topological magnon insulator, but it depends solely on the Berry curvature of the magnon bands.  We believe that  our result  offers the manipulation of the intrinsic property of 2D ferromagnetic insulators without an external magnetic field. This could pave the way for  investigating interesting potential practical applications of magnons in magnetic insulators,  such as  photo-magnonics \cite{benj}, magnon spintronics \cite{magn, benja},  and ultrafast optical control of magnetic spin currents \cite{ment, tak4, tak4a,walo}. 

\section*{ACKNOWLEDGEMNTS}
 Research at Perimeter Institute is supported by the Government of Canada through Industry Canada and by the Province of Ontario through the Ministry of Research
and Innovation.

\section*{ Appendix. Dirac magnons in an external electromagnetic field}

A massless charge-neutral Dirac magnon particle in the presence of external electromagnetic field  can be described  by the Dirac-Pauli Lagrangian \cite{paul,ho, bjo} 
\begin{align}
\mathcal L=\bar\psi\big(-v_0\gamma^0 + iv_D\gamma^\mu\partial_\mu-\frac{v_D\mu_m}{2}\sigma^{\mu\nu} F_{\mu\nu}\big)\psi,
\end{align}
where $v_0$ accounts for the finite energy Dirac magnon point and $v_D$ is the group velocity of the Dirac magnon. The two-component wave function is given by $\psi=(u_A, u_B)$  with $\bar\psi=\psi^\dg\gamma^0$. Here  $ F_{\mu\nu}$ is the electromagnetic field tensor and $\sigma^{\mu\nu}=\frac{i}{2}[\gamma^\mu,\gamma^\nu]=i\gamma^\mu\gamma^\nu,\quad (\mu\neq \nu).$

In (2+1) dimensions,   the Dirac matrices  are simply Pauli matrices given by
\begin{align}
\beta=\gamma^0=\sigma_z,~\gamma^1=i\sigma_y,~\gamma^2=-i\sigma_x.
\end{align}
They  obey the algebra \bea \lbrace \gamma^\mu,\gamma^\nu\rbrace=2g^{\mu\nu},
\eea 
 where $ g^{\mu\nu}=\text{diag}(1,-1,-1)$ is the Minkowski metric in 2+1 dimensions. The corresponding Hamiltonian is given by
 \begin{align}
 H = \int d^2 x ~\big [ \pi(x)\dot{\Psi}(x) - \mathcal L \big ] \equiv \int d^2 x ~ \Psi^\dg \mathcal H_D \Psi,
 \end{align}
 where $\pi(x) = \frac{\partial \mathcal L}{\partial{\dot{\Psi}(x)}}$ 
is the generalized momentum. 

We consider an electromagnetic field tensor with only time-varying electric field component  $\vec{E}(\tau)$. In this case, the corresponding Hamiltonian is given by
\begin{align}
\mathcal H_D=v_0 + v_D\vec{\sigma}\cdot\big[-i\vec{\nabla}+\mu_m\big(\vec{E}(\tau)\times \hat z\big)\big].
\end{align}
We can see that the time-dependent Aharonov-Casher phase  enters the Dirac magnon Hamiltonian as a minimal coupling.


\begin{thebibliography}{99}

\bibitem{mag}
J. Fransson, A. M. Black-Schaffer, and A. V. Balatsky, Phys. Rev. B {\bf 94}, 075401 (2016).
\bibitem{sol}
S. A.  Owerre, J. Phys.: Condens. Matter {\bf 28}, 386001 (2016).
\bibitem{kkim}
S. K. Kim et al., Phys. Rev. Lett. {\bf 117}, 227201 (2016).
\bibitem{pie}
Pierre A. Pantale\'on and Y. Xian, J. Phys.: Condens. Matter {\bf 29}, 295701 (2017).
\bibitem{per}
S. S. Pershoguba et al., Phys. Rev. X, {\bf 8} 011010 (2018).
\bibitem{boy}
D. Boyko, A. V. Balatsky, and J. T. Haraldsen, Phys. Rev. B {\bf 97}, 014433 (2018).
\bibitem{yago}
Y. Ferreiros, and Mar\'ia A. H. Vozmediano,  Phys. Rev. B {\bf 97}, 054404 (2018).
\bibitem{ruc}
A. R\"{u}ckriegel, A. Brataas, and R. A. Duine,  Phys. Rev. B {\bf 97}, 081106 (2018).
\bibitem{cas}
A. H. Castro Neto et al., Rev. Mod. Phys. {\bf 81}, 109 (2009).

\bibitem{fdm}
F. D. M. Haldane, Phys. Rev. Lett. {\bf 61}, 2015 (1988).
\bibitem{kane}
C. L. Kane and E. J. Mele, Phys. Rev. Lett. {\bf 95}, 146802 (2005).
\bibitem{Huang}
B. Huang et al., Nature (London) {\bf 546}, 270 (2017).

\bibitem{Gong}
C. Gong et al., Nature (London) {\bf 546}, 265 (2017).
 \bibitem{dm}
 I. Dzyaloshinsky, J. Phys. Chem. Solids {\bf 4}, 241 (1958).
  \bibitem{dm2}
   T. Moriya, Phys. Rev. {\bf 120}, 91 (1960).



\bibitem{aha}
 Y. Aharonov  and A. Casher,  Phys. Rev. Lett. {\bf 53}, 319 (1984).
  \bibitem{aha1}
Y. Aharonov  and D. Bohm, Phys. Rev. {\bf 115}, 485 (1959).
 \bibitem{ahaz}
 Z. Cao, X. Yu, and R. Han, Phys. Rev. B. {\bf 56}, 5077 (1997).
  \bibitem{ahat}
T. Liu and G. Vignale, Phys. Rev. Lett. {\bf 106}, 247203 (2011).
  \bibitem{loss}
F. Meier and  D. Loss, Phys. Rev. Lett. {\bf 90}, 167204 (2003).
 \bibitem{tak4}
  X.  Zhang  et al.,   Phys. Rev. Lett. {\bf 113}, 037202 (2014). 
\bibitem{mei}
 K. Nakata, J. Klinovaja,  and D. Loss,  Phys. Rev. B. {\bf 95}, 125429 (2017).
 
\bibitem{owe1}
S. A. Owerre, J. Phys. Commun. {\bf 1}, 021002 (2017).


\bibitem{foot3}
T. Oka,  and  H.  Aoki, Phys. Rev. B {\bf 79}, 081406 (2009).

\bibitem{foot5}
J. -I. Inoue and A. Tanaka, Phys. Rev. Lett. {\bf 105}, 017401 (2010).
\bibitem{foot4}
T. Kitagawa et al., Phys. Rev. B {\bf 84}, 235108 (2011).
\bibitem{fot1}
J. Cayssol et al.,  Physica Status Solidi (RRL)-Rapid Research Letters {\bf 7}, 101 (2013).
\bibitem{tp6}
P. Delplace,  \'A.  G\'omez-Le\'on, and Gloria Platero, Phys. Rev. B {\bf 88}, 245422 (2013).

\bibitem{tp7}
\'A. G\'omez-Le\'on, P. Delplace, and G. Platero, Phys. Rev. B {\bf 89}, 205408 (2014).
\bibitem{tp8}
A. G. Grushin, \'A.  G\'omez-Le\'on, and T. Neupert, Phys. Rev. Lett. {\bf 112}, 156801 (2014).

\bibitem{jot}
 G. Jotzu  et al.  Nature {\bf 515}, 237 (2014).
   \bibitem{eck1}
 A. Narayan, Phys. Rev. B {\bf 91}, 205445 (2015).
 \bibitem{eck2}
 Kush Saha, Phys. Rev. B {\bf 94}, 081103(R) (2016).
 \bibitem{eck3}
 Q. Chen, L. Du, and G. A. Fiete, Phys. Rev. B {\bf 97}, 035422 (2018).
 
 \bibitem{paul}
 W. Pauli, Rev. Mod. Phys. {\bf 13}, 203 (1941).
\bibitem{bjo} 
J. D. Bjorken  and S. D. Drell, \textit{Relativistic Quantum Mechanics}  (New York McGraw-Hill) (1964).

\bibitem{fuk}
T. Fukui, Y. Hatsugai and H. Suzuki, J. Phys. Soc. Jpn. {\bf 74}, 1674 (2005).

\bibitem{thm4}
 H. Katsura, N. Nagaosa, and P. A. Lee,   Phys. Rev. Lett.  {\bf 104},  066403 (2010).
     \bibitem{thm2}
Y. Onose et al.,  Science  { \bf 329}, 297 (2010).
  \bibitem{thm5}
 R. Matsumoto and S. Murakami, Phys. Rev. Lett. {\bf 106}, 197202 (2011); Phys. Rev. B. {\bf 84}, 184406 (2011).


 \bibitem{thm3}
T. Ideue et al., Phys. Rev. B. {\bf 85}, 134411 (2012).
\bibitem{thm6}
A.  Mook, J.  Henk, and I. Mertig, Phys. Rev. B {\bf 90}, 024412 (2014);  {\bf 89}, 134409 (2014).
 \bibitem{thm}
M. Hirschberger et al.,   Phys. Rev. Lett. {\bf 115}, 106603 (2015).
\bibitem{thm1}
R. Chisnell et al.,  Phys. Rev. Lett. {\bf 115}, 147201  (2015).
 \bibitem{sol1}
S. A.  Owerre, J. Appl. Phys. {\bf 120}, 043903 (2016).
\bibitem{foot2}
J. L. Lado and  J. F. -Rossier,   2D Mater. {\bf 4},  035002 (2017).
\bibitem{ste}
S. M. Young and C. L. Kane, Phys. Rev. Lett.  {\bf 115}, 126803 (2015).




\bibitem{benj}
   B. Lenk et al., arXiv:1208.5383 (2012).
 \bibitem{magn}
A. V.  Chumak    et al.,  Nat. Phys. {\bf 11}, 453 (2015).
  
\bibitem{benja}
 B. Lenk   et al.,  Phys. Rep. {\bf 507}, 107 (2011). 
 
  \bibitem{ment}
J. H. Mentink, J. Phys.: Condens. Matter {\bf 29} 453001 (2017).

 
  \bibitem{tak4a}
A. J.  Schellekens   et al.,  Nat. Commun. {\bf 5}, 4333 (2014).
 \bibitem{walo}
J. Walowski and M. M\"{u}nzenberg,    J. Appl. Phys. {\bf 120}, 140901 (2016). 

\bibitem{ho}
C. -L. Ho and P. Roy, Ann. Phys. {\bf 312},  161 (2004).

\end{thebibliography}
\end{document}